\documentclass{webofc}
\usepackage[varg]{txfonts}   % Web of Conferences font
\begin{document}
\title{Generation of baryon asymmetry in the E$_6$CHM}

\author{\firstname{Roman} \lastname{Nevzorov}\inst{1}\fnsep\thanks{\email{nevzorov@itep.ru}} \and
        \firstname{Anthony} \lastname{Thomas}\inst{2}
%        \fnsep\thanks{\email{Mail address for second
%             author if necessary}} \and
%       \firstname{Third author} \lastname{Third author}\inst{3}\fnsep\thanks{\email{Mail address for last
%             author if necessary}}
        % etc.
}

\institute{Alikhanov Institute for Theoretical and Experimental Physics, Moscow 117218, Russia
\and
ARC Centre of Excellence for Particle Physics at the Terascale and CSSM,
Department of Physics, The University of Adelaide, Adelaide SA 5005, Australia
%\and
%           Last address
          }

\abstract{%
The strongly interacting sector in the $E_6$ inspired composite Higgs model (E$_6$CHM)
with baryon number violation possesses an $\mbox{SU(6)}\times U(1)_L$ global symmetry.
In the weakly-coupled sector of this model the $U(1)_L$ symmetry 
associated with lepton number conservation
is broken down to a $Z^L_2$ discrete symmetry, which stabilizes the proton. Near the scale 
$f\gtrsim 10\,\mbox{TeV}$ the $\mbox{SU(6)}$ symmetry is 
broken down to its $\mbox{SU(5)}$ subgroup,
giving rise to a set of pseudo-Nambu-Goldstone bosons (pNGBs) that involves the SM-like Higgs
doublet, a scalar coloured triplet and a SM singlet boson. Because $f$ is so high, all baryon number 
violating operators are sufficiently strongly suppressed. Nevertheless, 
in this variant of the E$_6$CHM the observed matter–antimatter asymmetry can be induced 
if CP is violated. The pNGB scalar coloured triplet plays a key role in this process and leads to
a distinct signature that may be detected at the LHC in the near future.
}
\maketitle
\section{Introduction}
\label{intro}
The presence of baryon asymmetry and dark matter in the Universe provides
strong evidence for physics beyond the Standard Model (SM). Over the last forty years
a number of new physics mechanisms for baryogenesis have been proposed, including GUT baryogenesis \cite{gut-baryogen-1,gut-baryogen-2,gut-baryogen-3,gut-baryogen-4,gut-baryogen-5,gut-baryogen-6,gut-baryogen-7}, baryogenesis via leptogenesis \cite{Fukugita:1986hr}, the Affleck-Dine mechanism
\cite{Affleck-Dine-1, Affleck-Dine-2}, electroweak baryogenesis \cite{ew-baryogen}, etc\,.
Here we study baryon asymmetry generation within the $E_6$ inspired composite Higgs model (E$_6$CHM).

Composite Higgs models involve two sectors (for a review, see Ref.~\cite{Bellazzini:2014yua}).
One of them contains weakly-coupled elementary particles with the quantum numbers of all SM gauge bosons
and SM fermions. The second, strongly interacting sector results in a set of bound states that,
in particular, include the Higgs doublet and massive fields with the quantum numbers of all SM particles.
These massive fields are associated with the composite partners of the quarks, leptons and gauge bosons.
In composite Higgs models the elementary states couple to the composite operators of the strongly interacting
sector, leading to mixing between these states and their composite partners. Therefore at low energies
those states associated with the SM fermions (bosons) are superpositions of the corresponding elementary
fermion (boson) states and their composite fermion (boson) partners. In this framework, which is called
partial compositeness, the SM fields couple to the composite states, including the Higgs boson, with
a strength which is proportional to the compositeness fraction of each SM field. The observed mass
hierarchy in the quark and lepton sectors can be accommodated through partial compositeness if
the fractions of compositeness of the first and second generation fermions are rather small.
In this case the off-diagonal flavour transitions, as well as modifications of the $W$ and $Z$ couplings
associated with these light quark and lepton fields are somewhat suppressed. On the other hand, 
the top quark can be heavy if the right-handed top quark $t^c$ has a sizeable fraction of compositeness.

In the minimal composite Higgs model (MCHM) \cite{Agashe:2004rs} the strongly interacting sector possesses
a global $\mbox{SO(5)}\times U(1)_X$ symmetry that contains the $\mbox{SU(2)}_W\times U(1)_Y$ subgroup.
Near the scale $f$ the $\mbox{SO(5)}$ symmetry is broken down to $\mbox{SO(4)}$ so that
the SM gauge group remains intact. Such a breakdown of the global $\mbox{SO(5)}$ symmetry gives rise
to four pseudo-Nambu-Goldstone bosons (pNGBs) which form the Higgs doublet.
The couplings of the weakly-coupled elementary states to the strongly interacting
sector explicitly break the $\mbox{SO(5)}$ global symmetry. As a result,
the pNGB Higgs potential arises from loops involving elementary states. Because of this the quartic Higgs
coupling $\lambda$ in the Higgs effective potential is suppressed. The global custodial symmetry
$\mbox{SU(2)}_{cust} \subset \mbox{SU(2)}_W\times \mbox{SU(2)}_R$, which remains intact, 
protects the Peskin-Takeuchi $\hat{T}$ parameter against new physics 
contributions. Experimental limits on the value of the
parameter $|\hat{S}|\lesssim 0.002$ imply that $m_{\rho}=g_{\rho} f \gtrsim 2.5\,\mbox{TeV}$,
where $m_{\rho}$ is a scale associated with the masses of the set of spin-1 resonances and 
$g_{\rho}$ is a coupling of these $\rho$-like vector resonances. This set of resonances, 
in particular, involves composite partners of the SM gauge bosons.

While in composite Higgs models partial compositeness substantially reduces the contributions
of composite states to dangerous flavour-changing processes, this suppression is not sufficient
to satisfy all constraints. Adequate suppression of the non--diagonal flavour transitions can
be obtained only if $f$ is larger than $10\,\mbox{TeV}$. This bound on the scale $f$ can be considerably
alleviated in the composite Higgs models with additional flavour symmetries $\mbox{FS}$.
In models with $\mbox{FS}=U(2)^3=U(2)_{q}\times U(2)_u \times U(2)_d$ symmetry, the
constraints originating from the Kaon and $B$ systems can be satisfied even for $m_{\rho}\sim 3\,\mbox{TeV}$.
For low values of the scale $f$ the appropriate suppression of the baryon number violating operators
and the Majorana masses of the left-handed neutrino can be achieved provided global $U(1)_B$ and
$U(1)_L$ symmetries, which guarantee the conservation of baryon and lepton numbers, respectively,
are imposed.
\section{E$_6$CHM with baryon number violation}
\label{sec-1}
In the $E_6$ inspired composite Higgs model (E$_6$CHM) the Lagrangian of the strongly interacting
sector is invariant under the transformations of an $\mbox{SU(6)}\times U(1)_L$ global symmetry.
This model can be embedded into $N=1$ supersymmetric (SUSY) orbifold Grand Unified Theories (GUTs)
in six dimensions which are based on the $E_6\times G_0$ gauge group \cite{e6chm}. At some high
energy scale, $M_X$, the $E_6\times G_0$ gauge symmetry is broken down to the
$\mbox{SU(3)}_C\times \mbox{SU(2)}_W\times U(1)_Y \times G$ subgroup where
$\mbox{SU(3)}_C\times \mbox{SU(2)}_W\times U(1)_Y$ is the SM gauge group.
The gauge groups $G_0$ and $G$ are associated with the strongly coupled sector.
Multiplets from this sector can be charged under both the $E_6$ and $G_0$ ($G$) gauge symmetries.
The weakly--coupled sector comprises elementary states that participate in the $E_6$ interactions only.
It is expected that in this sector different multiplets of the elementary quarks and leptons
are components of different bulk $27$--plets, whereas all other components of these $27$--plets
should acquire masses of the order of $M_X$. The appropriate splitting of the $27$--plets can occur
within the six--dimensional orbifold GUT model with $N=1$ supersymmetry (SUSY) \cite{e6chm}
in which SUSY is broken near the GUT scale $M_X$. (Different phenomenological aspects of the
$E_6$ inspired models with low-scale SUSY breaking were recently examined in \cite{King:2005jy}-\cite{King:2016wep}.).

In this orbifold GUT model all fields from the strongly interacting sector reside
on the brane where $E_6$ symmetry is broken down to the $\mbox{SU(6)}$ that contains an
$\mbox{SU(3)}_C\times \mbox{SU(2)}_W\times U(1)_Y$ subgroup. Thus at high energies
the Lagrangian of the composite sector respects $\mbox{SU(6)}$ symmetry.
The SM gauge interactions violate this global symmetry.  Nevertheless, $\mbox{SU(6)}$
can remain an approximate global symmetry of the strongly interacting sector at low energies
if the gauge couplings of this sector are substantially larger than the SM ones.
Around the scale $f$ the $\mbox{SU(6)}$ symmetry in the E$_6$CHM is expected to be broken
down to its $\mbox{SU(5)}$ subgroup, so that the SM gauge group is preserved.
Since E$_6$CHM does not possesses any extra custodial or flavour symmetry,
the scale $f$ should be larger than $10\,\mbox{TeV}$. Such a breakdown of the $\mbox{SU(6)}$
global symmetry gives rise to a set of pNGB states, which includes the SM--like Higgs doublet.

In the E$_6$CHM the $U(1)_L$ global symmetry, which ensures the conservation of lepton number,
suppresses the operators in the strongly coupled sector that induce too large Majorana masses
of the left--handed neutrino. In the weakly--coupled elementary sector this symmetry
should be broken down to
\begin{equation}
Z^L_{2}=(-1)^{L} \, ,
\label{1}
\end{equation}
where $L$ is a lepton number, to guarantee that the left--handed neutrinos gain small
Majorana masses. The $Z^L_2$ discrete symmetry, which is almost exact, forbids operators that
give rise to rapid proton decay.

All other baryon number violating operators are sufficiently
strongly suppressed by the large value of the scale $f$, as well as the rather small mixing
between elementary states and their composite partners. Indeed, the operators responsible for
$\Delta B=2$ and $\Delta L=0$ are given by
\begin{equation}
\mathcal{L}_{\Delta B=2}=\frac{1}{\Lambda^5}\Biggl[ q_i q_j q_k q_m (d^c_n d^c_l)^{*} +
u^c_i d^c_j d^c_k u^c_m d^c_n d^c_l \Biggr]\,,
\label{2}
\end{equation}
where $q_i$ are doublets of left-handed quarks, $u^c_i$ and $d^c_j$ are the right-handed
up- and down-type quarks and $i,j,k,m,n,l=1,2,3$.
The $n-\bar{n}$ mixing mass is $\delta m \simeq \varkappa \Lambda_{QCD}^6/ \Lambda^5$,
where $\varkappa\sim 1$ and $\Lambda_{QCD}\simeq 200\,\mbox{MeV}$.
For $\Lambda \sim \mbox{few}\times 100\,\mbox{TeV}$ one finds the free $n-\bar{n}$ oscillation time
to be $\tau_{n-\bar{n}}\simeq 1/\delta m \simeq 10^8\,\mbox{s}$. This value of $\tau_{n-\bar{n}}$
is quite close to the present experimental limit \cite{Phillips:2014fgb}.
Searches for rare nuclear decays caused by the annihilation of the two nucleons $NN\to KK$, that
can be also induced by the operators (\ref{2}), result in a similar lower limit on $\Lambda$, 
around $200-300\,\mbox{TeV}$. At the same time the small mixing between elementary states
and their composite partners in the composite Higgs models leads to
$\Lambda\gtrsim \mbox{few}\times 100\,\mbox{TeV}$ when $f\gtrsim 10\,\mbox{TeV}$.

Although the $U(1)_{B}$ global symmetry associated with baryon number conservation
is not imposed here we assume that the low energy effective Lagrangian of the E$_6$CHM
is invariant with respect to an approximate $Z^B_2$ symmetry,
which is a discrete subgroup of $U(1)_{B}$, i.e.
\begin{equation}
Z^B_{2}=(-1)^{3B} \, ,
\label{3}
\end{equation}
where $B$ is the baryon number. The $Z^B_2$ discrete symmetry does not forbid
baryon number violating operators (\ref{2}). Nevertheless it provides an additional mechanism
for the suppression of proton decay.

The embedding of the E$_6$CHM into a Grand Unified Theory (GUT) based on
the $E_6\times G_0$ gauge group implies that the SM gauge couplings extrapolated to
high energies using the renormalisation group equations (RGEs) should converge to some
common value near the scale $M_X$. In the E$_6$CHM an approximate unification of
the SM gauge couplings can be achieved if the right--handed top quark $t^c$ is entirely
composite and the sector of weakly--coupled elementary states involves  \cite{e6chm}
\begin{equation}
(q_i,\,d^c_i,\,\ell_i,\,e^c_i) + u^c_{\alpha} + \bar{q}+\bar{d^c}+\bar{\ell}+\bar{e^c}\,,
\label{4}
\end{equation}
where $\alpha=1,2$ and $i=1,2,3$. In Eq.~(\ref{4}) we have denoted the left-handed lepton
doublet by $\ell_i$ and the right-handed charged leptons by $e_i^c$, while the extra exotic
states $\bar{q},\,\bar{d^c},\,\bar{\ell}$ and $\bar{e^c}$, have exactly opposite
$\mbox{SU(3)}_C\times \mbox{SU(2)}_W\times U(1)_Y$ quantum numbers to the left-handed quark doublets,
right-handed down-type quarks, left-handed lepton doublets and right-handed charged leptons,
respectively. The set of elementary fermion states (\ref{4}) is chosen so that
the weakly--coupled sector involves all SM fermions except the right--handed top quark
and anomaly cancellation takes place.

Using the one--loop RGEs one can find the value of $\alpha_3(M_Z)$ for which 
exact gauge coupling unification takes place in the E$_6$CHM
\begin{equation}
\dfrac{1}{\alpha_3(M_Z)}=\dfrac{1}{b_1-b_2}\biggl[\dfrac{b_1-b_3}{\alpha_2(M_Z)}-
\dfrac{b_2-b_3}{\alpha_1(M_Z)}\biggr]\,,
\label{41}
\end{equation}
where $b_i$ are one--loop beta functions, with the indices $1,\,2,\,3$ corresponding
to the $U(1)_Y$, $\mbox{SU(2)}_W$ and $\mbox{SU(3)}_C$ interactions. Since all composite states form complete
$\mbox{SU(5)}$ multiplets, the strongly interacting sector does not contribute to the differential
running determined by $(b_i-b_j)$ in the one--loop approximation. Then, it is rather easy
to find that, for $\alpha(M_Z)=1/127.9$ and $\sin^2\theta_W=0.231$, exact gauge coupling
unification within the E$_6$CHM takes place near $M_X\sim 10^{15}-10^{16}\, \mbox{GeV}$,
with $\alpha_3(M_Z)\simeq 0.109$\,. This estimate indicates that for
$\alpha_3(M_Z)\simeq 0.118$ the SM gauge couplings can be reasonably close to each other
at very high energies around $M_X\simeq 10^{16}\, \mbox{GeV}$.

The E$_6$CHM implies that the dynamics of the strongly coupled sector gives rise to the composite
${\bf 10} + {\bf \overline{5}}$ multiplets of $\mbox{SU(5)}$. These multiplets get combined with $\bar{q},\,\bar{d^c},\,\bar{\ell}$ and $\bar{e^c}$, forming a set of vector--like states.
The only exceptions are the components of the $10$--plet that correspond to $t^c$,
which survive down to the electroweak (EW) scale. In the simplest scenario the composite
${\bf 10} + {\bf \overline{5}}$ multiplets of $\mbox{SU(5)}$ may stem from one ${\bf{15}}$--plet and
two ${\bf \overline{6}}$--plets (${\bf \overline{6}}_1$ and ${\bf \overline{6}}_2$) of $\mbox{SU(6)}$.
These $\mbox{SU(6)}$ representations decompose under $\mbox{SU(3)}_C\times \mbox{SU(2)}_W\times U(1)_Y$ as follows:
\begin{equation}
\begin{array}{ll}
\begin{array}{rcl}
{\bf 15} &\to& Q = \left(3,\,2,\,\dfrac{1}{6}\right)\,,\\[0mm]
&& t^c = \left(3^{*},\,1,\,-\dfrac{2}{3}\right)\,,\\[0mm]
&& E^c = \Biggl(1,\,1,\,1\Biggr)\,,\\[0mm]
&& D = \left(3,\,1,\,-\dfrac{1}{3} \right)\,,\\[0mm]
&& \overline{L}=\left(1,\,2,\,\dfrac{1}{2}\right)\,;
\end{array}
\qquad
\noindent
\begin{array}{rcl}
{\bf \overline{6}}_{\alpha} &\to & D^c_{\alpha} = \left(\bar{3},\,1,\,\dfrac{1}{3} \right)\,,\\[0mm]
& & L_{\alpha} = \left(1,\,2,\,-\dfrac{1}{2} \right)\,,\\[0mm]
& & N_{\alpha} = \Biggl(1,\,1,\,0 \Biggr)\,,
\end{array}
\end{array}
\label{5}
\end{equation}
where $\alpha=1,2$. In Eq.~(\ref{5}) the first and second quantities in brackets are the $\mbox{SU(3)}_C$
and $\mbox{SU(2)}_W$ representations, while the third ones are the $U(1)_Y$ charges. Because $t^c$ is
$Z^B_2$-odd, all components of the ${\bf{15}}$--plet should be odd under the $Z^B_2$ symmetry.
After the breakdown of the $\mbox{SU(6)}$ symmetry a ${\bf 5}$--plet from the ${\bf{15}}$--plet
and ${\bf \overline{5}}$--plet from the ${\bf \overline{6}}_2$ compose vector--like states.
This means that all components of ${\bf \overline{6}}_2$ should be $Z^B_2$-odd.
Hereafter the components of ${\bf \overline{6}}_1$ multiplet are assumed to be $Z^B_2$--even.

In general all fermion states mentioned above gain masses which are larger than $f$.
Here we assume that $N_1$ is considerably lighter than other fermion states and has a mass which
is somewhat smaller than $f$. The mixing between the SM singlet states $N_1$ and $N_2$
should be suppressed because of the approximate $Z^B_2$ symmetry.
\section{Baryon asymmetry generation}
The breakdown of the $\mbox{SU(6)}$ symmetry to its $\mbox{SU(5)}$ subgroup near the
scale $f\gtrsim 10\,\mbox{TeV}$ gives rise to eleven pNGB states that can be parameterised by
\begin{equation}
\begin{array}{c}
\Omega^T = \Omega_0^T \Sigma^T = e^{i\frac{\phi_0}{\sqrt{15}f}}
\Biggl(C \phi_1\quad C \phi_2\quad C \phi_3\quad C \phi_4\quad C\phi_5\quad \cos\dfrac{\tilde{\phi}}{\sqrt{2} f} + \sqrt{\dfrac{3}{10}} C \phi_0 \Biggr)\,,\\[3mm]
C=\dfrac{i}{\tilde{\phi}} \sin \dfrac{\tilde{\phi}}{\sqrt{2} f}\,,\qquad \tilde{\phi}=\sqrt{\dfrac{3}{10}\phi_0^2+|\phi_1|^2+|\phi_2|^2+|\phi_3|^2+|\phi_4|^2+|\phi_5|^2}\,,
\end{array}
\label{51}
\end{equation}
$$
\Omega_0^T= (0\quad 0\quad 0\quad 0\quad 0\quad 1)\,,\qquad \Sigma= e^{i\Pi/f}\,,\qquad
\Pi=\Pi^{\hat{a}} T^{\hat{a}}\,.
$$
where $T^{\hat{a}}$ are broken generators of $\mbox{SU(6)}$. The masses of all pNGB states tend to be
considerably lower than $f$. Thus these resonances are the lightest composite states in the E$_6$CHM.
The set of pNGB states involves one real SM singlet scalar, $\phi_0$, one $\mbox{SU(2)}_W$ doublet
$H\sim (\phi_1\, \phi_2)$, that corresponds to the SM--like Higgs doublet, as well as
$\mbox{SU(3)}_C$ triplet of scalar fields $T\sim (\phi_3\,\, \phi_4\,\, \phi_5)$. The collider signatures
associated with the presence of the SM singlet scalar, $\phi_0$, were examined in
Ref.~\cite{Nevzorov:2016fxp, quarks-2016}.

The Lagrangian, that describes the interactions of the pNGB states, in the leading approximation
can be written as
\begin{equation}
\mathcal{L}_{pNGB}=\frac{f^2}{2}\biggl|\mathcal{D}_{\mu} \Omega \biggr|^2\,.
\label{52}
\end{equation}
The pNGB effective potential $V_{eff}(H, T, \phi_0)$ can be obtained by integrating out
the heavy resonances of the strongly coupled sector. It is induced by the interactions of the
elementary fermions and gauge bosons with their composite partners that violate the $\mbox{SU(6)}$ symmetry.
In the E$_6$CHM substantial tuning, at a level $\sim 0.01\%$, is needed to get $v\ll f$ and
a $125\,\mbox{GeV}$ Higgs boson. In the framework of models similar to the E$_6$CHM
it was demonstrated that there exists a part of the parameter space where the $\mbox{SU(2)}_W\times U(1)_Y$
gauge symmetry is broken to $U(1)_{em}$, whilst $\mbox{SU(3)}_C$ remains intact
\cite{Frigerio:2011zg, Barnard:2014tla}. In these models the $\mbox{SU(3)}_C$ triplet of scalars, $T$,
is substantially heavier than the SM--like Higgs boson.

Since the Higgs boson is a $Z^B_2$-even state, all other pNGB states should also be even under
the $Z^B_2$ symmetry. As a consequence, the $\mbox{SU(3)}_C$ scalar triplet can decay into
up and down antiquarks. Because the fractions of compositeness of the first and second
generation quarks are very small the decay mode $T\to\bar{t}\bar{b}$ has to be the dominant one.
At the energies $E\lesssim f$ all baryon number violating operators are strongly suppressed. Thus
baryon number is conserved to a very good approximation and $T$ manifests itself in the
interactions with other SM particles as a diquark, i.e. an $\mbox{SU(3)}_C$ scalar triplet with
$B=-2/3$. At the LHC, the $\mbox{SU(3)}_C$ scalar triplet can be pair produced resulting in the
enhancement of the cross section for the process $pp\to T\bar{T}\to t\bar{t}b\bar{b}$.

In the E$_6$CHM the baryon asymmetry can be generated via the out--of equilibrium
decays of $N_1$, provided $N_1$ is the lightest composite fermion in the spectrum.
Because $m_{T}\ll f$ the decays $N_{1}\to T+\bar{d}_i$ and $N_{1}\to T^{*}+ d_i$ are
allowed, resulting in final states with baryon numbers $\pm 1$. The Lagrangian,
that describes the interactions of $N_1$ and $N_2$ with $T$ and down-type quarks, is given by
\begin{equation}
\mathcal{L}_{N}= \sum_{i=1}^3 \Biggl( g^{*}_{i1} T d^c_i N_1 + g^{*}_{i2} T d^c_i N_2 + h.c.\Biggr) \, .
\label{6}
\end{equation}
In~ the~ exact~ $Z^B_2$~ symmetry~ limit,~ $g_{i1}=0$. The approximate $Z^B_2$ symmetry
ensures that $|g_{i1}| \ll |g_{i2}|$.

The process of the generation of the baryon asymmetry is controlled by the three
CP (decay) asymmetries 
\begin{equation}
\varepsilon_{1,\,k}=\dfrac{\Gamma_{N_1 d_{k}}-\Gamma_{N_1 \bar{d}_{k}}}
{\sum_{m} \left(\Gamma_{N_1 d_{m}}+\Gamma_{N_1 \bar{d}_{m}}\right)} \, ,
\label{7}
\end{equation}
where $\Gamma_{N_1 d_{k}}$ and $\Gamma_{N_1 \bar{d}_{k}}$ are partial decay widths 
of $N_1\to d_k + T^{*}$ and $N_1\to \overline{d}_k + T$ with $k,m=1,2,3$.
These decay asymmetries vanish at the tree level because \cite{bau-e6chm}
\begin{equation}
\Gamma_{N_1 d_{k}}=\Gamma_{N_1 \bar{d}_{k}}=\dfrac{3 |g_{k1}|^2}{32 \pi}\,M_1\,.
\label{8}
\end{equation}
In Eq.~(\ref{8}) $M_1$ is the Majorana mass of $N_1$. 

Nevertheless, non--zero contributions to the CP asymmetries may arise from the interference 
between the tree--level amplitudes of the $N_1$ decays and the one--loop corrections to them.
Since $T$ couples primarily to the third generation fermions, $|g_{31}|\gg |g_{21}|,\, |g_{11}|$ 
and $|g_{32}|\gg |g_{22}|,\, |g_{12}|$. This hierarchical structure of the Yukawa interactions
implies that the decay asymmetries $\varepsilon_{1,\,2}$ and $\varepsilon_{1,\,1}$ are much smaller 
than $\varepsilon_{1,\,3}$ and can be ignored in the leading approximation. Assuming that
$N_1$ is much lighter than other composite fermion states including $N_2$ and $m_{T}\ll M_1$,
so that $m_{T}$ can be neglected, the direct calculation of one--loop diagrams gives
\begin{equation}
\varepsilon_{1,\,3}\simeq -\dfrac{1}{(4\pi)}\dfrac{|g_{32}|^2}{\sqrt{x}}\sin 2\Delta\varphi\,,\qquad\qquad
\Delta\varphi=\varphi_{32}-\varphi_{31}\,,
\label{11}
\end{equation}
where $x=(M_2/M_1)^2$, $M_2$ is the Majorana mass of $N_2$, $g_{31}=|g_{31}| e^{i\varphi_{31}}$
and $g_{32}=|g_{32}| e^{i\varphi_{32}}$. The decay asymmetry (\ref{11}) vanishes if CP invariance 
is preserved, i.e. all Yukawa couplings are real. The absolute value of the CP asymmetry 
$|\varepsilon_{1,\,3}|$ attains its maximum when $\Delta\varphi=\pm \pi/4$.

In order to calculate the total baryon asymmetry induced by the decays of $N_1$, one needs to
solve the system of Boltzmann equations that describe the evolution of baryon number densities. 
Because the corresponding solution has to be similar to the solutions of the Boltzmann equations 
for thermal leptogenesis, the generated baryon asymmetry can be estimated using an approximate 
formula \cite{bau-e6chm}
\begin{equation}
Y_{\Delta B}\sim 10^{-3}\biggl(\varepsilon_{1,\,3} \eta_3\biggr)\,,
\label{10}
\end{equation}
where $\eta_3$ is an efficiency factor, that varies from 0 to 1, and 
$Y_{\Delta B}$ is the baryon asymmetry relative to the entropy density, i.e.
\begin{equation}
Y_{\Delta B}=\dfrac{n_B-n_{\bar{B}}}{s}\biggl|_0=(8.75\pm 0.23)\times 10^{-11}\,.
\label{9}
\end{equation}
A thermal population of $N_1$ decaying completely out of equilibrium without washout 
effects would result in $\eta_{3}=1$. However, washout processes
reduce the generated baryon asymmetry by the factor $\eta_3$. The induced baryon 
asymmetry should be partially converted into lepton asymmetry due to $(B+L)$--violating 
sphaleron interactions \cite{Kuzmin:1985mm, sphaleron}. Here we ignore these processes.

In the strong washout scenario the efficiency factor $\eta_3$ can be estimated as follows 
\begin{equation}
\begin{array}{c}
\eta_3 \simeq H(T=M_1)/\Gamma_{3}\,,\\[3mm]
\Gamma_3 = \Gamma_{N_1 d_{3}}+\Gamma_{N_1 \bar{d}_{3}}=\dfrac{3 |g_{31}|^2}{16 \pi}\,M_1\,,
\qquad\qquad
H=1.66 g_{*}^{1/2}\dfrac{T^2}{M_{Pl}}\,,
\end{array}
\label{12}
\end{equation}
where $H$ is the Hubble expansion rate and $g_{*}=n_b+\dfrac{7}{8}\,n_f$ is the number of relativistic
degrees of freedom in the thermal bath. In the SM $g_{*}=106.75$, while in the E$_6$CHM
$g_{*}=113.75$ for $T\lesssim f$. Eqs.~(\ref{12}) indicate that the efficiency factor $\eta_3$ increases 
with diminishing of $|g_{31}|$. For very small values of this coupling $\eta_3$ becomes close to unity. 
For example, when $|g_{31}|\simeq 10^{-6}$ and $M_1\simeq 10\,\mbox{TeV}$ the efficiency factor 
$\eta_3\simeq 0.25$.

\begin{figure}[h]
\centering
\includegraphics[width=10cm,clip]{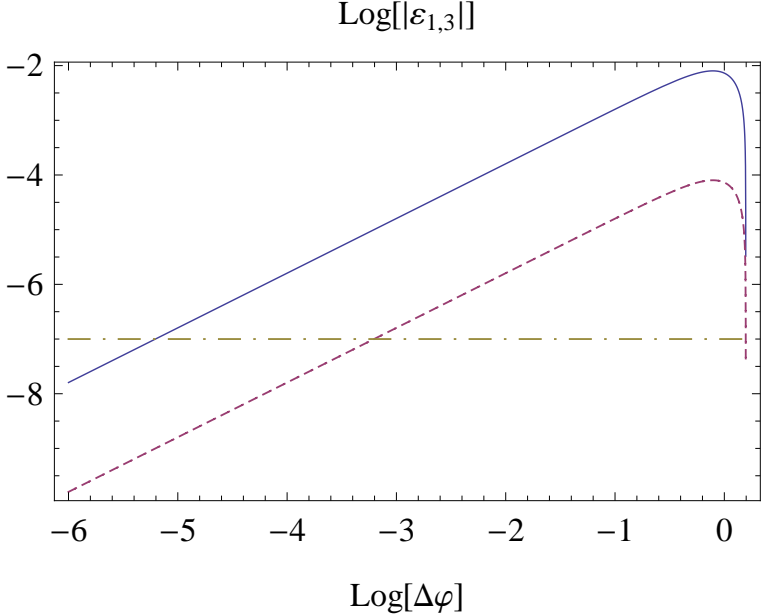}
\caption{Logarithm (base 10) of the absolute value of the CP asymmetry $\varepsilon_{1,\,3}$ 
as a function of logarithm (base 10) of $\Delta \varphi$ for $|g_{32}|=1$ (solid line) and 
$|g_{32}|=0.1$ (dashed line) in the case when $g_{11}=g_{21}=g_{12}=g_{22}=0$
and $M_2=10\cdot M_1$.}
\label{fig-quarks-18}
\end{figure}

When $\eta_3 \sim 0.1-1$, the induced baryon asymmetry is determined by the CP asymmetry 
$\varepsilon_{1,\,3}$. From Eqs.~(\ref{11}) it follows that for a given ratio $M_2/M_1$ 
the decay asymmetry $\varepsilon_{1,\,3}$ is set by $|g_{32}|$ and the combination of
the CP--violating phases $\Delta \varphi$ but does not depend on the absolute value of 
the Yukawa coupling $g_{31}$. Since the Yukawa coupling of $N_2$ to $\mbox{SU(3)}_C$ scalar triplet $T$
and $b$-quark is not suppressed by the $Z_2^B$ symmetry, $|g_{32}|$ can be relatively 
large, i.e. $|g_{32}| \gtrsim 0.1$. In Fig.~\ref{fig-quarks-18} we explore the dependence 
of $|\varepsilon_{1,\,3}|$ on $\Delta \varphi$ for $|g_{32}| = 0.1$ and $|g_{32}| = 1$.
We fix $(M_2/M_1)=10$. The maximum absolute value of $\varepsilon_{1,\,3}$ grows monotonically 
with increasing of $|g_{32}|$. The results presented in Fig.~\ref{fig-quarks-18} demonstrate that
the decay asymmetry, $\varepsilon_{1,\,3}$, attains its maximum absolute value, $\sim 10^{-4}-10^{-2}$, 
for $\Delta \varphi\simeq \pi/4$. For such large values of $|\varepsilon_{1,\,3}|$ a phenomenologically 
acceptable baryon density can be obtained only if $\eta_3=10^{-5}-10^{-3}$. When $\eta_3 \sim 0.1-1$
and $|g_{32}|\simeq 0.1$ a phenomenologically acceptable value of the baryon density,~ corresponding~ 
to~ $\varepsilon_{1,\,3}\lesssim10^{-7}-10^{-6}$~ can~ be~ induced~ if~ $\Delta \varphi$~ is~ rather~ small,~ 
i.e.~ $\Delta\varphi\lesssim 0.01$. Thus the appropriate baryon asymmetry can be generated within the 
E$_6$CHM even in the case when CP is approximately preserved.  

\begin{acknowledgement}
A.W.T. was supported by the University of Adelaide and the
Australian Research Council through the ARC Center of Excellence
in Particle Physics at the Terascale (CE 110001004).
\end{acknowledgement}

\end{document}